\documentclass[prc,aps,twocolumn,showpacs,amssymb,superscriptaddress]{revtex4}

\usepackage{amsmath}
\usepackage{graphicx}
\usepackage{dcolumn}
\usepackage{float}
\begin{document}

\title{Double-step truncation procedure for large-scale shell-model calculations}

\author{L. Coraggio}
\affiliation{Istituto Nazionale di Fisica Nucleare, \\
Complesso Universitario di Monte  S. Angelo, Via Cintia - I-80126 Napoli, Italy}
\author{A. Gargano}
\affiliation{Istituto Nazionale di Fisica Nucleare, \\
Complesso Universitario di Monte  S. Angelo, Via Cintia - I-80126 Napoli, Italy}
\author{N. Itaco}
\affiliation{Istituto Nazionale di Fisica Nucleare, \\ 
Complesso Universitario di Monte  S. Angelo, Via Cintia - I-80126 Napoli, Italy}
\affiliation{Dipartimento di Matematica e Fisica, Seconda Universit\`a
di Napoli, viale Abramo Lincoln 5 - I-81100 Caserta, Italy}

\begin{abstract}
We present a procedure that is helpful to reduce the computational
complexity of large-scale shell-model calculations, by preserving as
much as possible the role of the rejected degrees of freedom in
an effective approach.
Our truncation is driven first by the analysis of the effective
single-particle energies of the original large-scale shell-model
hamiltonian, so to locate the relevant degrees of freedom to describe
a class of isotopes or isotones, namely the single-particle orbitals
that will constitute a new truncated model space.
The second step is to perform an unitary transformation of the
original hamiltonian from its model space into the truncated one.
This transformation generates a new shell-model hamiltonian, defined
in a smaller model space, that retains effectively the role of the
excluded single-particle orbitals.
As an application of this procedure, we have chosen a realistic
shell-model hamiltonian defined in a large model space, set up by
seven and five proton and neutron single-particle orbitals outside
$^{88}$Sr, respectively.
We study the dependence of shell-model results upon different
truncations of the original model space for the Zr, Mo, Ru, Pd, Cd,
and Sn isotopic chains, showing the reliability of this truncation
procedure.
\end{abstract}


\maketitle

\section{Introduction}
\label{intro}
Large-scale shell-model (LSSM) calculations have become a well-established
approach to obtain a microscopic theoretical description of the collective
properties of atomic nuclei.
This goal has been evident since the early papers where the rotational
motion of some $pf$-shell nuclei \cite{Caurier94} or the large
$\gamma$-softeness of $^{64}$Ge \cite{Honma96}, already observed in
experimental studies, were described in terms of the shell model (SM)
instead of resorting to nuclear collective models.

In present days, powerful computing devices are widely accessible and
make more feasible to approach SM calculations with large model spaces
for nuclei with many valence nucleons.
This has given the opportunity to many nuclear-theory groups to study
exotic features of the atomic nuclei within a microscopic approach, so
supporting the experimental efforts to enlarge the knowledge of the
chart of the nuclides in the rare-ion-beam era.
In recent years, it is worth mentioning the study of the onset of
collectivity in some isotopic and isotonic chains
\cite{Gade10,Ljungvall10,Baugher12,Crawford13}, the revelation of
novel collective features \cite{Lenzi10,Sieja13,Tsunoda14,Togashi15}
and of the island-of-inversion phenomenon \cite{Caurier14}, the
description of the shell evolution in many heavy-mass nuclei
\cite{Sieja10,Bianco13,Sahin15,Wang15}.
Along with these SM studies, it should be also pointed out that a
notable theoretical effort has been devoted to enhance the
computational abilities of LSSM calculations
\cite{Hasegawa01,Mizusaki12,Shimizu12,Johnson13,Utsuno14,Stumpf16}.

In spite of the progress in solving eigenproblems of large
complexity, there exists always a limit to the maximum dimension of
matrices that can be diagonalized when solving the Schr{\"o}dinger
equation.
This affects mostly the shell-model studies of heavy-mass nuclei,
where the required model spaces own a large capacity, making the
calculations for nuclei with many valence nucleons very demanding.

For example, in Ref. \cite{Banu05} experimental data for light tin
isotopes are compared with the results of LSSM calculations performed
by considering $^{90}$Zr as a closed core, and using the ANTOINE
shell-model code \cite{antoine}.
Since the model space was made up by the six $sdgh$ proton orbitals and
the five neutron ones (excluding the $0g_{9/2}$ orbital), exact
calculations were not feasible, and the SM basis was truncated allowing
up to $4p-4h$ $Z=50$ cross-shell excitations only.

A more recent example, which may be mentioned, is the observation of the
onset of collectivity at $N=40$ in the chromium and iron isotopic
chains \cite{Crawford13}.
The latter has been interpreted as due to the interplay between the
quadrupole-quadrupole component of the residual interaction and the
central field in the sub-space spanned by the lowest $\Delta j$=2
orbitals of a major shell \cite{Zuker95}.
This interpretation is based on a LSSM calculation, where the model
space is composed by the four $fp$ proton orbitals and five $fpgd$
neutron ones, with 4 and 6 valence protons and up to 12 valence
neutrons.
Also in this calculation, the diagonalization of the full SM basis,
using the NATHAN code \cite{NATHAN}, was not feasible, and a
truncation up to= $14p-14h$ excitations across the $Z=28$ and $N=40$
shell-closures was employed.

From the above examples, it is evident that computational
difficulties, when dealing with large model spaces, arise evolving
the number of the valence protons $Z_{val}$ (isotonic chains) and/or
of the valence neutrons $N_{val}$ (isotopic chains), a typical
situation for nuclei exhibiting exotic features.

The increase of the valence particles, however, comes along with the
evolution of the theoretical effective single-particle energies (ESPE)
for a given SM hamiltonian.
Therefore, the behavior of the ESPE as a function of the number of
valence protons or neutrons may help to identify the relevant degrees of
freedom to describe the spectroscopic characteristics of a certain nucleus.

In the present work, we propose a method, that we have already
experienced in Ref. \cite{Coraggio15a}, which locates a very
effective truncation of the model space, through the study of the ESPE
of a SM hamiltonian $H$ as a function of $Z_{val}$ and/or $N_{val}$.
Successively, we build up a new SM hamiltonian $\tilde{H}$ defined in
a reduced model space with a smaller number of orbitals, by way of an
unitary transformation of the ``mother hamiltonian'' $H$.

As a testing ground, we consider the study of the quadrupole collectivity
due to $Z=50$ cross-shell excitations in even-even isotopic chains
above $^{88}$Sr.

First, we derive an effective shell-model hamiltonian starting from the
CD-Bonn potential \cite{Machleidt01b}, whose high-momentum repulsive
components are smoothed out using the $V_{\rm low-k}$ approach
\cite{Bogner02}, by way of the time-dependent perturbation theory
\cite{Kuo71,Coraggio09a}.
This will be done within a large model space, that includes seven
$psdgh$ proton orbitals and five $sdgh$ neutron ones.

Then, we calculate and study the behavior of the proton and neutron
ESPE as a function of the number of valence neutrons and protons.
The analysis of the ESPE drives the choice to reduce the number of
proton and neutron orbitals.
Consequently, new effective hamiltonians defined in reduced
model spaces and tailored to study certain isotopic chains, are
derived by way of a unitary transformation of the starting shell-model
hamiltonian.
Finally, we perform SM calculations with these effective
hamiltonians and compare the theoretical results so to quantify the
reliability of this double-step procedure.

In the next section, we sketch out a few details about the derivation
of our shell-model hamiltonians and effective charges of the electric
quadrupole operators.
We also present the results of the behavior of the calculated ESPE as
a function of the valence protons.
Besides this, we will discuss how we have derived the new effective
hamiltonians, within two new truncated model spaces. 
In Section III, we present the results of our calculations for
Zr, Mo, Ru, Pd, Cd, and Sn isotopes, comparing the results obtained
with three different model spaces.
In the last section a summary of the present work and an outlook of
our future programs are reported.
In Ref. \cite{supplemental2016} they can be found the calculated
two-body matrix elements (TBME) of the effective shell-model
interactions in the truncated model spaces.

\section{Outline of calculations}
\label{calculations}
Our starting effective shell-model hamiltonian has been derived within the
framework of the many-body perturbation theory , as mentioned in the
Introduction, from the CD-Bonn $NN$ potential \cite{Machleidt01b}.
More explicitly, first the high-momentum repulsive components of the
bare $NN$ potential have been renormalized by way of the so-called $V_{\rm
  low-k}$ approach \cite{Bogner01,Bogner02}, which provides a smooth
potential preserving exactly the onshell properties of the original
$NN$ potential up to a cutoff momentum $\Lambda=2.6$ fm$^{-1}$.
Next, we have derived the shell-model hamiltonian using the well-known
$\hat{Q}$-box plus folded-diagram method \cite{Coraggio09a}, where the
$\hat{Q}$-box is a collection of irreducible valence-linked Goldstone
diagrams which we have calculated through third order in the $V_{\rm
  low-k}$ \cite{Coraggio12a}.

The effective hamiltonian $H_{\rm eff}$ can be written in an operator
form as 

\begin{equation}
H_{\rm eff} = \hat{Q} - \hat{Q'} \int \hat{Q} + \hat{Q'} \int \hat{Q} \int
\hat{Q} - \hat{Q'} \int \hat{Q} \int \hat{Q} \int \hat{Q} + ~...~~,
\end{equation}

\noindent
where the integral sign represents a generalized folding operation, 
and $\hat{Q'}$ is obtained from $\hat{Q}$ by removing terms of first
order in $V_{\rm low-k}$.
The folded-diagram series is summed up to all orders using the
Lee-Suzuki iteration method \cite{Suzuki80}.

\begin{table}[H]
\caption{Theoretical shell-model SP energy spacings (in MeV)
  employed in present work (see text for details).}
\begin{ruledtabular}
\begin{tabular}{ccc}
$nlj$ & proton SP energies & neutron SP energies \\
\colrule
 $1p_{1/2}$   & 0.0   &   ~   \\ 
 $0g_{9/2}$   & 1.5   &   ~   \\ 
 $0g_{7/2}$   & 5.7   &  1.5  \\ 
 $1d_{5/2}$   & 6.4   &  0.0  \\ 
 $1d_{3/2}$   & 8.8   &  3.4  \\ 
 $2s_{1/2}$   & 8.7   &  2.2  \\ 
 $0h_{11/2}$ & 10.2 &  5.1  \\ 
\end{tabular}
\end{ruledtabular}
\label{spetab}
\end{table}

\begin{table}[H]
\caption{Proton and neutron effective charges of the electric
  quadrupole operator $E2$.}
\begin{ruledtabular}
\begin{tabular}{ccc}
$n_a l_a j_a ~ n_b l_b j_b $ &  $\langle a || e_p || b \rangle $ &  $\langle a || e_n || b \rangle $ \\
\colrule
$0g_{9/2}~0g_{9/2}$     & 1.53  &   ~   \\ 
 $0g_{9/2}~0g_{7/2}$     & 1.58 &   ~   \\ 
 $0g_{9/2}~1d_{5/2}$     & 1.51 &   ~   \\ 
 $0g_{7/2}~0g_{9/2}$     & 1.77 &   ~  \\ 
 $0g_{7/2}~0g_{7/2}$     & 1.84 & 1.00 \\ 
 $0g_{7/2}~1d_{5/2}$     & 1.84 & 0.98 \\ 
 $0g_{7/2}~1d_{3/2}$     & 1.86 & 0.98 \\ 
 $1d_{5/2}~0g_{9/2}$     & 1.59 &  ~  \\ 
 $1d_{5/2}~0g_{5/2}$     & 1.73 & 0.92 \\ 
 $1d_{5/2}~1d_{5/2}$     & 1.73 & 0.87 \\ 
 $1d_{5/2}~1d_{3/2}$     & 1.71 & 0.90 \\ 
 $1d_{5/2}~2s_{1/2}$     & 1.76 & 0.73 \\ 
 $1d_{3/2}~0g_{7/2}$     & 1.83 & 0.94 \\ 
 $1d_{3/2}~1d_{5/2}$     & 1.79 & 0.93 \\ 
 $1d_{3/2}~1d_{3/2}$     & 1.81 & 0.92 \\ 
 $1d_{3/2}~2s_{1/2}$     & 1.83 & 0.75 \\ 
 $2s_{1/2}~1d_{5/2}$     & 1.73 & 0.73 \\ 
 $2s_{1/2}~1d_{3/2}$     & 1.73 & 0.73 \\ 
 $0h_{11/2}~0h_{11/2}$  & 1.89 & 0.87 \\ 
\end{tabular}
\end{ruledtabular}
\label{effch}
\end{table}

From the effective hamiltonian both single-particle (SP) energies and
two-body matrix elements of the residual interaction have been
obtained \cite{Coraggio12a}, and we have derived consistently the
effective charges of the electric quadrupole operators at the same
perturbative order.

The model space we have employed is spanned by the seven
$1p_{1/2},0g_{9/2},0g_{7/2},1d_{5/2},1d_{3/2},2s_{1/2},0h_{11/2}$
proton orbitals and by the five
$0g_{7/2},1d_{5/2},1d_{3/2},2s_{1/2},0h_{11/2}$ neutron orbitals,
considering $^{88}$Sr as an inert core.
From now on, we will indicate this model space as $[75]$, and
consequently, we have dubbed the effective hamiltonian $H^{75}$, the
superscript indicating the number of proton (seven) and neutron (five)
model-space orbitals.

Such a large model space has been introduced so as to
take explicitly into account the $Z=50$ cross-shell excitations of protons
jumping from the $1p_{1/2},0g_{9/2}$ orbitals into the $sdgh$ ones.

In Table \ref{spetab} the calculated single-particle energies are
reported, and in Table \ref{effch} are reported theoretical proton and
neutron effective charges, which are close to the usual empirical
values ($e^{\rm emp}_p=1.5e,~e^{\rm emp}_n =0.5\div 0.8e$).

As previously mentioned, the major difficulty with $H^{75}$
is the increasing computational complexity, when evolving the atomic
number $Z$ of the isotopic chain under investigation.
In particular, if one focuses his attention on the tin chain - as we
did in Ref. \cite{Coraggio15a} - this hamiltonian cannot be
diagonalized for any tin isotope with up-to-date shell-model codes.

It is then mandatory to find the way to reduce the dimensions of the
matrices to be diagonalized, and consequently make the shell-model
calculation feasible.

\begin{figure}[ht]
\begin{center}
\includegraphics[scale=0.38,angle=0]{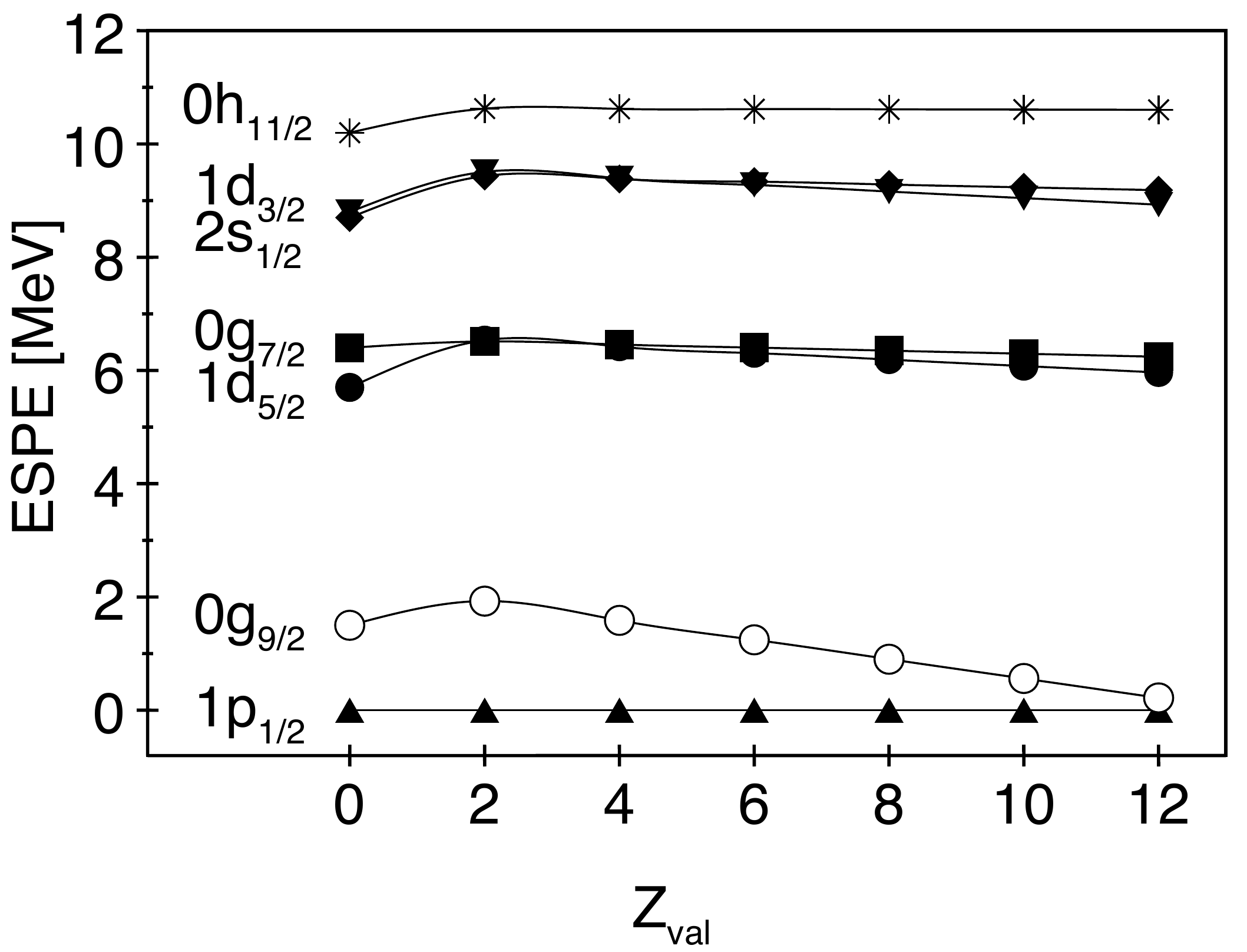}
\caption{Calculated proton effective single-particle energies of
  $H^{75}$ as a function of the number of valence protons $Z_{val}$.}
\label{espeZ}
\end{center}
\end{figure}

\begin{figure}[ht]
\begin{center}
\includegraphics[scale=0.38,angle=0]{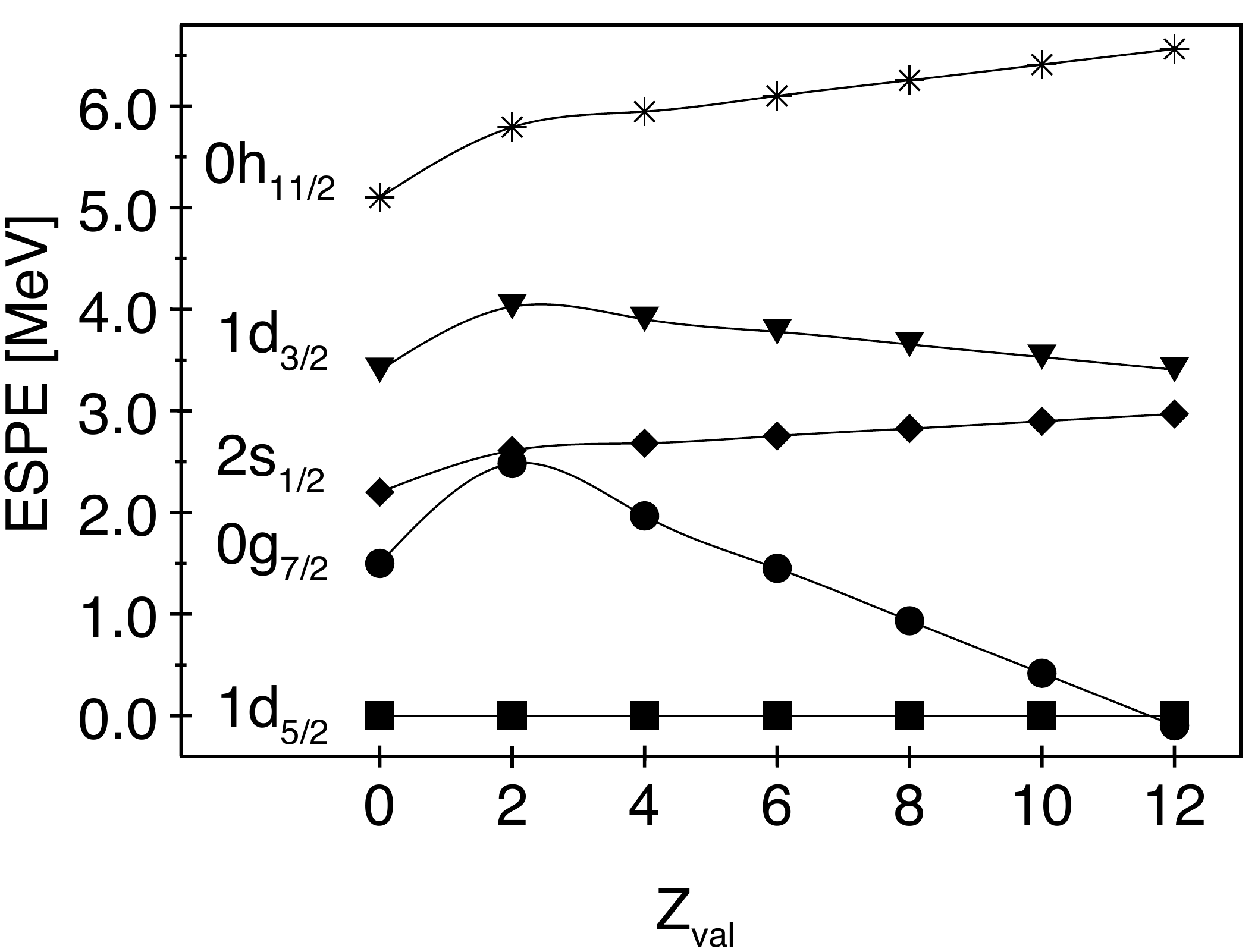}
\caption{Calculated neutron effective single-particle energies of
  $H^{75}$ as a function of the number of valence protons $Z_{val}$.}
\label{espeN}
\end{center}
\end{figure}

Here we present an approach, adopted for the first time in
Ref. \cite{Coraggio15a}, which leads to new effective hamiltonians
defined in truncated model spaces, by way of a unitary transformation
of $H^{75}$.
The choice of the truncation of the model space is driven by the
behavior, as a function of $Z_{val}$ and $N_{val}$, of the proton and neutron
effective single-particle energies (ESPE) of the original hamiltonian
$H^{75}$, so as to find out what are the most relevant
degrees of freedom to describe the physics of nuclear systems of interest.
To this end, we report in Figs. \ref{espeZ} and \ref{espeN} the evolution
of both proton and neutron ESPE as a function of $Z_{val}$.
From the inspection of Fig. \ref{espeZ}, it can be observed that an
almost constant energy gap provides a separation between the subspace
spanned by the $1p_{1/2},0g_{9/2},1d_{5/2},0g_{7/2}$ proton orbitals
and that spanned by the $2s_{1/2},1d_{3/2},0h_{11/2}$ ones.
This leads to the conclusion that a reasonable truncation is to
consider only the lowest four orbitals, as proton model space.\

On the neutron side, Fig. \ref{espeN} evidences that the filling of the
proton $0g_{9/2}$ orbital induces a relevant energy gap at $Z=50$
between the $1d_{5/2},0g_{7/2}$ subspace and that spanned by the
$2s_{1/2},1d_{3/2},0h_{11/2}$ orbitals.
This gap, around 2.4 MeV, traces back to the tensor component of the
proton-neutron interaction that is mainly responsible for the shell
evolution \cite{Otsuka05}.

On the above grounds, it looks reasonable that a neutron model space
spanned only by the $1d_{5/2},0g_{7/2}$ orbitals may provide the
relevant features of the physics of isotopic chains with higher $Z$,
such as Pd, Cd, and Sn isotopes.

Consequently, we have derived two new effective hamiltonians
$H_{\rm eff}^{45},H_{\rm eff}^{42}$, defined within two model spaces
$[45],[42]$ spanned by the $1p_{1/2},0g_{9/2},1d_{5/2},0g_{7/2}$
proton orbitals, and by the
$0g_{7/2},1d_{5/2},1d_{3/2},2s_{1/2},0h_{11/2}$ and
$0g_{7/2},1d_{5/2}$ neutron orbitals, respectively.

Here, we present a few details about the derivation of these effective
hamiltonians, dubbed $H_{\rm eff}^{pn}$, starting from the ``mother
hamiltonian'' $H^{75}$.

The eigenvalue problem for $H^{75}$ written in terms of its
eigenvalues $E_i$ and eigenfunctions $\psi_i$ is the following:
\begin{equation}
H^{75}| \psi_i \rangle= E_i | \psi_i \rangle ~~,
\end{equation}

\noindent
where $H^{75}$ may be partitioned as the sum of a single-particle
hamiltonian $H_0$ and a residual two-body potential $V$:
\begin{equation}
H^{75}=H_0 + V~~.
\end{equation}

As it has been mentioned before, the analysis of the behavior of the
ESPE induces a possible reduction of the number SP orbitals that span
the model space.
The original model space $[75]$ is then splitted up in two subspaces
$P \equiv P^{pn}$ and $Q \equiv Q^{7-p,5-n}$, with the projector $P$
that can be expressed in terms of the $H_0$ eigenvectors
\begin{equation}
P = \sum_{i=1,d} |i \rangle \langle i | ~~~~~~~~~~~~~~   H_0| i
\rangle= E^0_i | i \rangle ~~.
\end{equation} 

We can define the $P$-space effective hamiltonian $H_{\rm eff}^{pn}$ by writing
\begin{equation}
H_{\rm eff}^{pn}| \phi_k \rangle= \left( PH_oP + V_{\rm eff}^{pn} \right) | \phi_k \rangle
= E_k | \phi_k \rangle~~,
\label{h1}
\end{equation}

\noindent
where we require that the eigenfunctions $\phi_k$ are the projections of the
eigenfunctions $\psi_k$ of the ``mother hamiltonian'' 
\[
| \phi_k \rangle = P | \psi_k \rangle~~.
\]

Then we can express $H_{\rm eff}^{pn}$ formally as
\begin{equation}
H_{\rm eff}^{pn} = \sum_{k=1}^d E_k | \phi_k \rangle\langle \tilde{\phi_k}|~~,
\end{equation}

\noindent
where the $| \tilde{\phi_k} \rangle$ are the $| \phi_k \rangle$
biorthogonal states satisfying $| \tilde{\phi_k} \rangle \langle
\phi_{k'}|= \delta_{kk'}$, and that can be easily obtained using the
Schmidt biorthonormalization procedure. 

At the end, we can therefore define the effective residual interaction
$V_{\rm eff}^{pn}$ as: 
\begin{equation}
V_{\rm eff}^{pn} = \sum_{k=1}^d E_k | \phi_k \rangle\langle \tilde{\phi_k}| - PH_0P ~~.
\end{equation}

The knowledge of the eigenvalues and eigenfunctions of $H^{75}$ is therefore
essential to explicitly derive the effective hamiltonian $H_{\rm eff}^{pn}$.

It is worth to point out that when solving the $H^{75}$ eigenvalue problem
for a $A_{val}$ valence-nucleon system, the corresponding effective
hamiltonian $H_{\rm eff}^{pn}$ will contain 1-body, 2-body, ... $A_{val}$-body
contributions.
Obviously, the larger is the chosen subspace the smaller is
the role of these effective $A_{val}$-body components.

Since at present, due to the computational complexity, there are no
public shell-model codes able to handle these $A_{val}$-body forces with $n
\geq 3$, we have applied the above unitary transformation only to the
two valence-nucleon systems (i.e. $^{90}$Zr,$^{90}$Sr,$^{90}$Y), thus taking into
account only TBME of $H_{\rm eff}^{45},H_{\rm eff}^{42}$. 

This implies that the energy spectra of
$^{90}$Zr,$^{90}$Sr, and $^{90}$Y are exactly the same when
diagonalizing $H^{75}$, $H_{\rm eff}^{45}$, and $H_{\rm eff}^{42}$. 

\begin{widetext}
\begin{center}
\begin{figure}[H]
\hspace{2.0truecm}
\includegraphics[scale=0.60,angle=0]{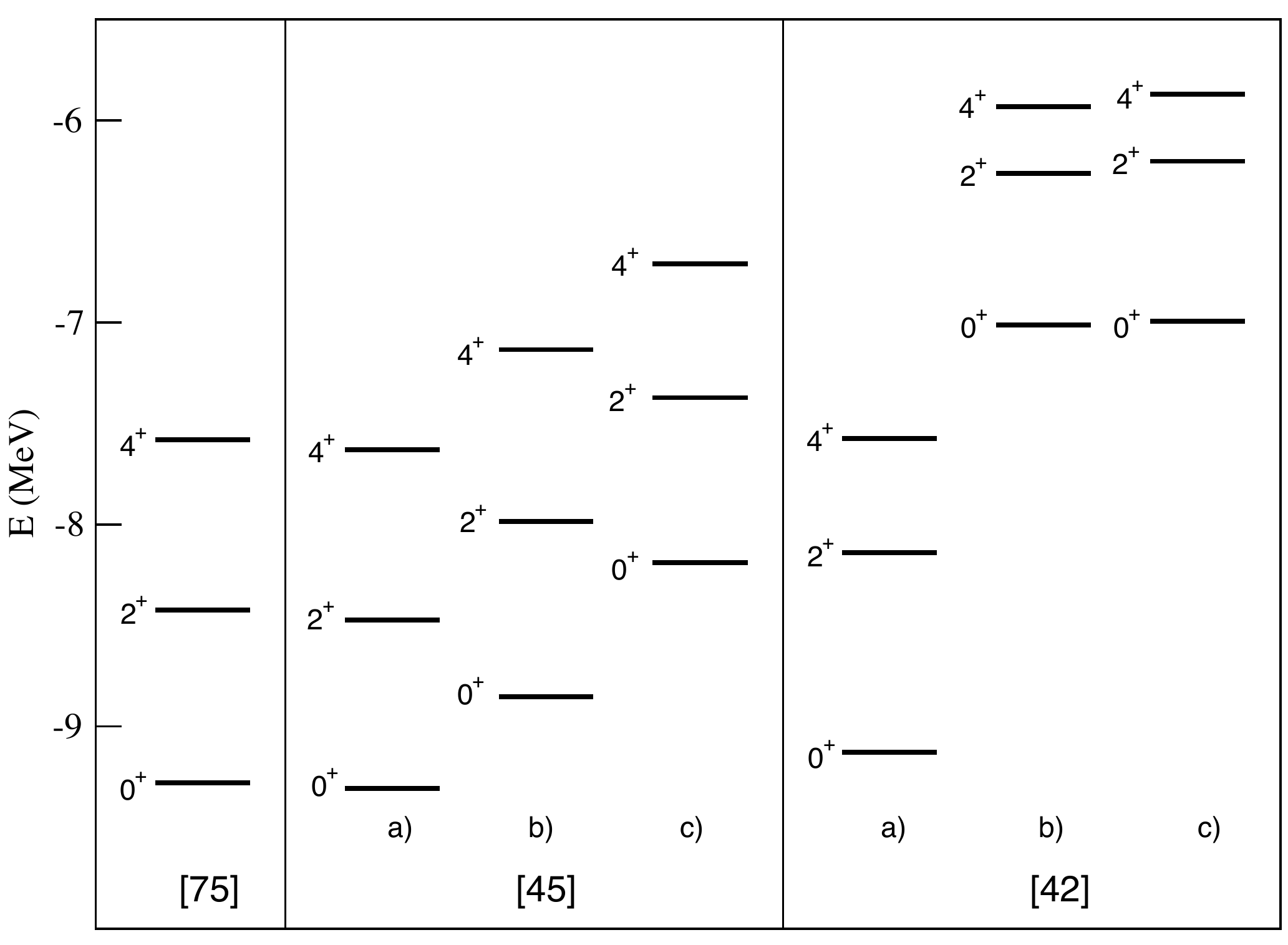}
\caption{Yrast $J=0^+,2^+,4^+$ states of $^{96}$Mo calculated within
  model spaces $[7,5]$, $[4,5]$, and $[4,2]$ (see text for details).}
\label{96Mo}
\end{figure}
\end{center}
\end{widetext}

To verify the reliability of our truncation scheme, we consider the
nucleus $^{96}$Mo, which, in our framework, is described as four
protons and four neutrons interacting outside $^{88}$Sr, and
corresponds to the ``largest'' eigenvalue problem solvable with the
``mother'' hamiltonian $H^{75}$ for a system with an equal
number of valence protons and neutrons.

The shell model calculations with the truncated model spaces $[4,5]$
and $[4,2]$ are performed employing:
\begin{enumerate}
\item[a)] the effective hamiltonians $H_{\rm eff}^{45}$ and $H_{\rm eff}^{42}$;
\item[b)] two effective hamiltonians $\tilde{H}_{\rm
      eff}^{45},\tilde{H}_{\rm eff}^{42}$ derived by way of the
  many-body perturbation theory within the model spaces $[4,5]$ and
  $[4,2]$;
\item[c)] the hamiltonian $H^{75}$, but constraining the calculations
  to model spaces $[45],[42]$.
\end{enumerate}

In Fig. \ref{96Mo} we compare the absolute energies of yrast $J=
0^+,2^+,4^+$ states in $^{96}$Mo obtained by means of the a),b),c)
shell-model calculations.
From the inspection of Fig. \ref{espeN}, we expect that the $H_{\rm
  eff}^{42}$ results will be in a not so good agreement with the
$H^{75}$ ones, since there is not a clear separation of the
model space $P$ from its complement $Q$. 

It can be noted from Fig. \ref{96Mo} that $H_{\rm eff}^{45}$ is able
to reproduce quite well the absolute energies of the ``mother
hamiltonian'' $H^{75}$, while this is not the case for
$H_{\rm eff}^{42}$.
More precisely, even if $H_{\rm eff}^{42}$ reproduces nicely the 2$^+$
excitation energy, it underestimates the collectivity predicted by the
``mother hamiltonian''.
As a matter of fact, the $R_{4/2}$ ratio between the calculated
excitation energies of the 4$^+$ versus the 2$^+$ states, that is
equal to 2.0 with $H^{75}$ and $H_{\rm eff}^{45}$, drops to
1.6 when evaluated with $H_{\rm eff}^{42}$. 
For the sake of completeness, it should be observed that the spectra
obtained employing the b) and c) hamiltonians are both shifted up with
respect to case a), the size of the shift being proportional to the
truncation of the model space.

The above results evidence the adequacy of our truncation scheme when
the latter is grounded on a neat separation of the model space
$P$ from its complement $Q$, as depicted by the ESPE behavior (see
Figs. \ref{espeZ},\ref{espeN}).

As mentioned before, we have calculated the $E2$ effective charges
consistently with the ``mother hamiltonian'' $H^{75}$ at the
same perturbative order.
This means that, when dealing with the effective hamiltonians $H_{\rm eff}^{pn}$,
in order to preserve exactly also the calculated transition rates for
the two-valence nucleon systems, the effective $E2$ operator should be
further renormalized to take into account the neglected degrees of
freedom.
In this way, one would obtain an effective two-body $E2$ operator to
be employed to calculate the electric quadrupole properties of the
systems with a number of valence nucleons greater than two.
At our knowledge, however, there's no public shell-model code able to
calculate transition rates driven by two-body transition operators,
therefore we have calculated the $E2$ transition rates using the
effective charges derived consistently with  $H^{75}$.
As a consequence, the eventual observed discrepancy between the $E2$
properties calculated with $H^{75}$ and those with the effective
hamiltonians $H_{\rm eff}^{pn}$ is always a signature of the fact that the
corresponding $H^{75}$ wave functions have relevant
components outside the truncated $[pn]$ model space. 

\section{Results}
We report the results we have obtained for the even $Z \leq 50$
isotopic chains, namely Zr, Mo, Ru, Pd, Cd, and Sn isotopes with even
neutron number.
We compare the results obtained using the three different effective
hamiltonians $H^{75}$, $H_{\rm eff}^{45}$, and $H_{\rm eff}^{42}$, focusing attention on
the quadrupole properties, i.e.  yrast $2^+$ excitation energies and
$B(E2;0^+_1 \rightarrow 2^+_1)$ transition rates.
Calculations have been performed using the public version of the
ANTOINE shell-model code \cite{antoine} .

\begin{figure}[H]
\begin{center}
\includegraphics[scale=0.37,angle=0]{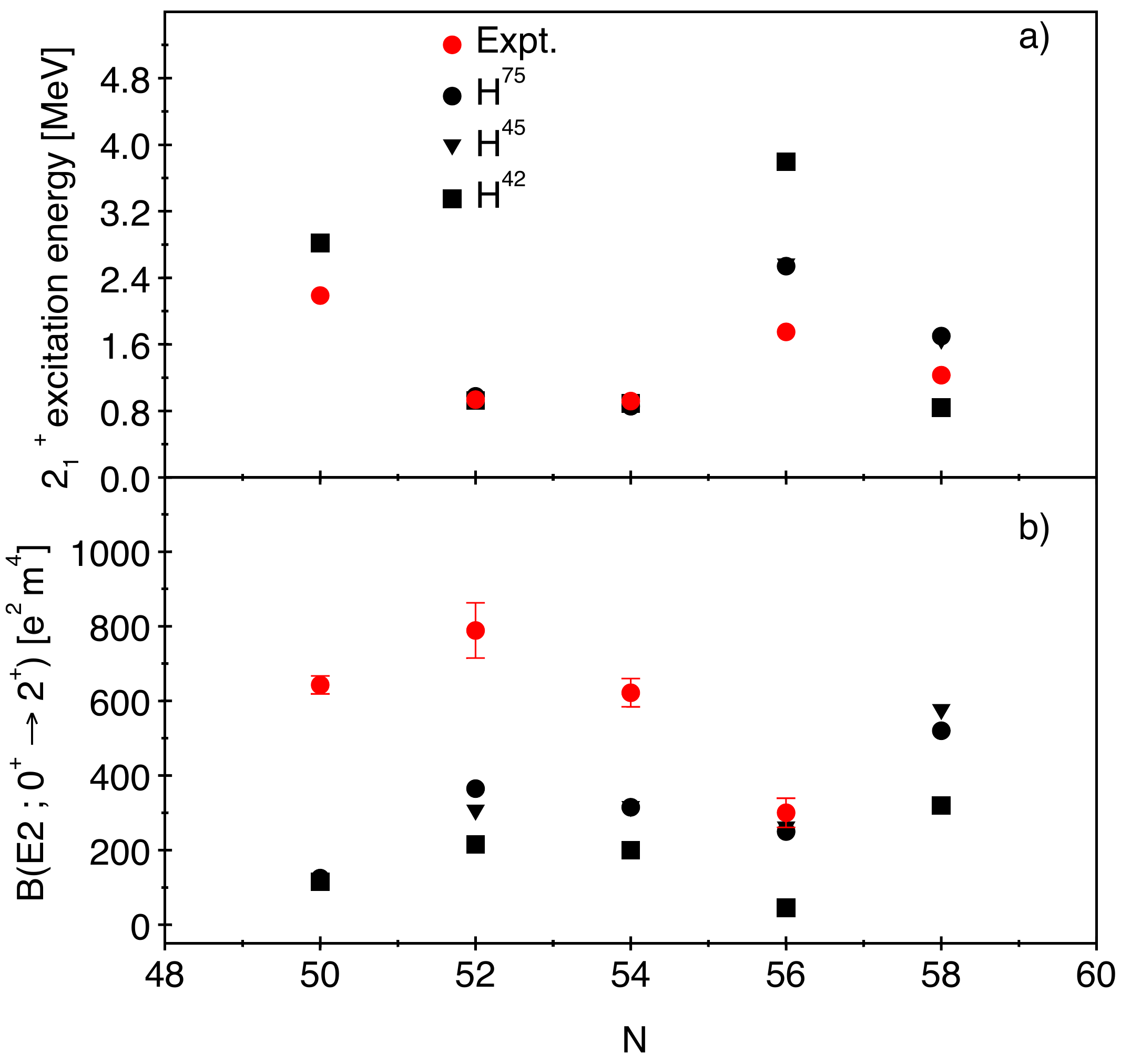}
\caption{(Color online) (a) Experimental \cite{ensdf}
  (red circles) and calculated (black symbols) excitation energies of the
  yrast $J^{\pi}=2^+$ states and (b) $B(E2;2^+_1 \rightarrow 0^+_1)$
  transition rates for zirconium isotopes as a function of $N$.  Circles,
  triangles, and squares refer to results obtained with $H^{75}$,
  $H_{\rm eff}^{45}$, and $H_{\rm eff}^{42}$, respectively.} 
\label{Zr}
\end{center}
\end{figure}

\subsection{Zr isotopes}
In our framework Zr isotopes are described as two valence protons and
$(N-50)$ valence neutrons outside $^{88}$Sr, therefore $^{90}$Zr is a
two-valence particle system and by definition, as mentioned in
Sec. \ref{calculations}, its calculated energy spectrum is exactly
preserved by the three effective hamiltonians.
This can be seen in Fig. \ref{Zr} where the calculated excitation
energies of the yrast $2^+$ states ($E^{ex}_{2^+_1}$) and the
$B(E2;0^+_1 \rightarrow 2^+_1)$ transition  rates are reported as a
function of $N$, and compared with the experimental data.
The calculated $E^{ex}_{2^+_1}$ are indistinguishable up to $N=54$,
while for $N=56, 58$ the $H_{\rm eff}^{42}$ eigenvalues are quite different from
the $H^{75}$ and $H_{\rm eff}^{45}$ ones, the former predicting a shell closure
for $N=56$.

\begin{figure}[H]
\begin{center}
\includegraphics[scale=0.37,angle=0]{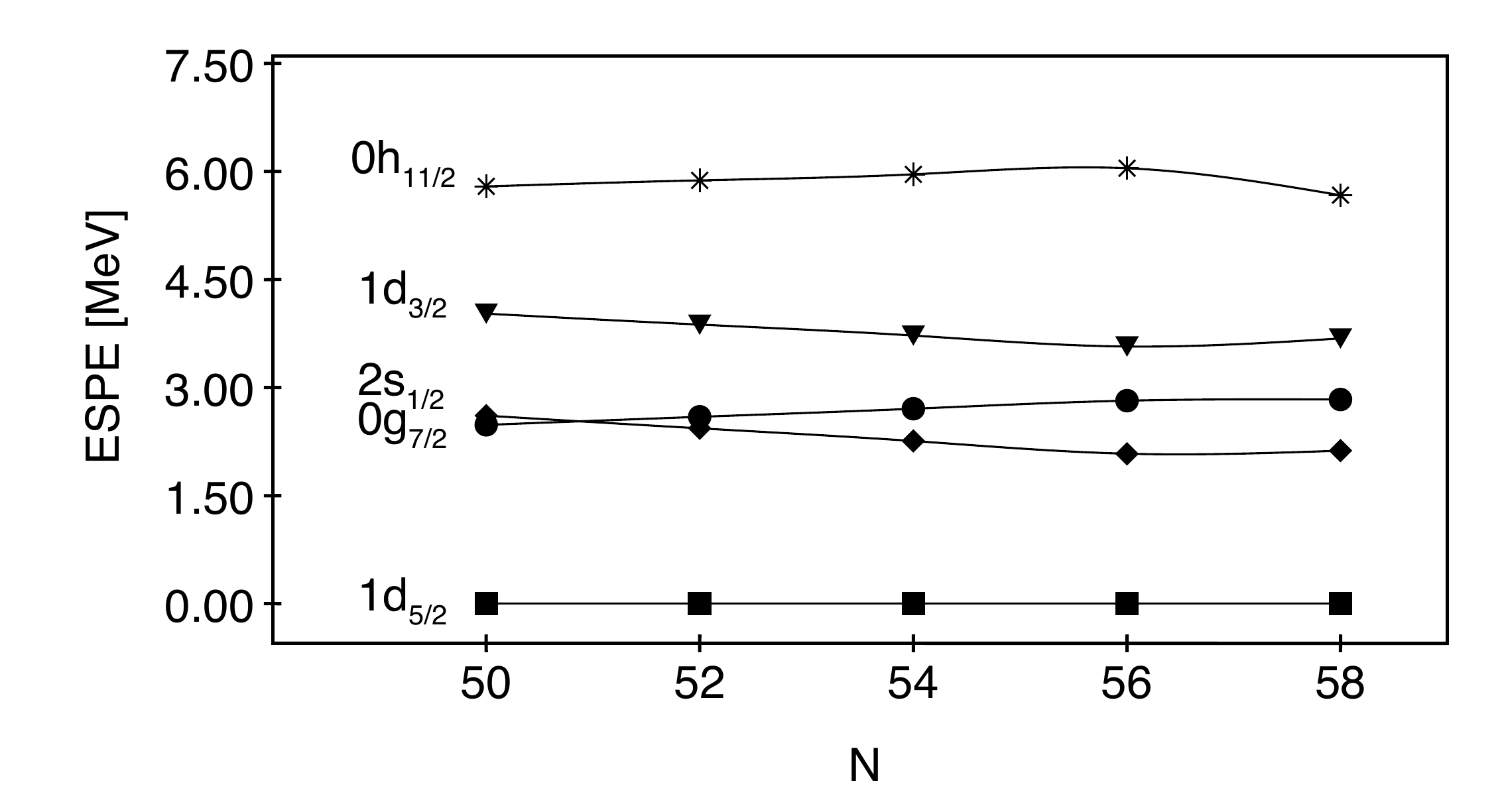}
\caption{Zr neutron ESPE calculated from $H^{75}$ as a function of  $N$.}
\label{espeZr}
\end{center}
\end{figure}

The discrepancy obtained using $H_{\rm eff}^{42}$, when evolving the number of
neutrons, can be interpreted in terms of the Zr neutron ESPE, whose
behavior as a function of $N$ is reported in Fig. \ref{espeZr}. 
The two neutron orbitals $0g_{7/2}$ and $2s_{1/2}$ are very close in
energy, so that, when filling the $1d_{5/2}$ orbital, the reduction of
the neutron model space to only two orbitals becomes too severe, and
$H_{\rm eff}^{42}$ is not able to take into account effectively the neglected
$2s_{1/2}$ degrees of freedom. 

Similar considerations can be drawn for the electric quadrupole
transition rates; the agreement between the $H_{\rm eff}^{42}$ results and the
$H^{75}$ and  $H_{\rm eff}^{45}$ ones can be considered satisfactory up to
$N=54$, while deteriorating for heavier isotopes. 

As regards the comparison with the experiment, the observed behavior
of the $E^{ex}_{2^+_1}$ is well reproduced by the $H^{75}$ and
$H_{\rm eff}^{45}$ hamiltonians.
This is not the case for the $B(E2;0^+_1 \rightarrow 2^+_1)$ rates,
their values being strongly underestimated by the calculations. 
Our results are not able to describe adequately the quadrupole
collectivity that is observed for $N=50-54$. 

This may be traced back to the non-negligible role of the $Z=38$
proton cross-shell excitations, as pointed out in Ref. \cite{Sieja09},
that should be explicitly included in the calculations.

\subsection{Mo isotopes}
\begin{figure}[H]
\begin{center}
\includegraphics[scale=0.37,angle=0]{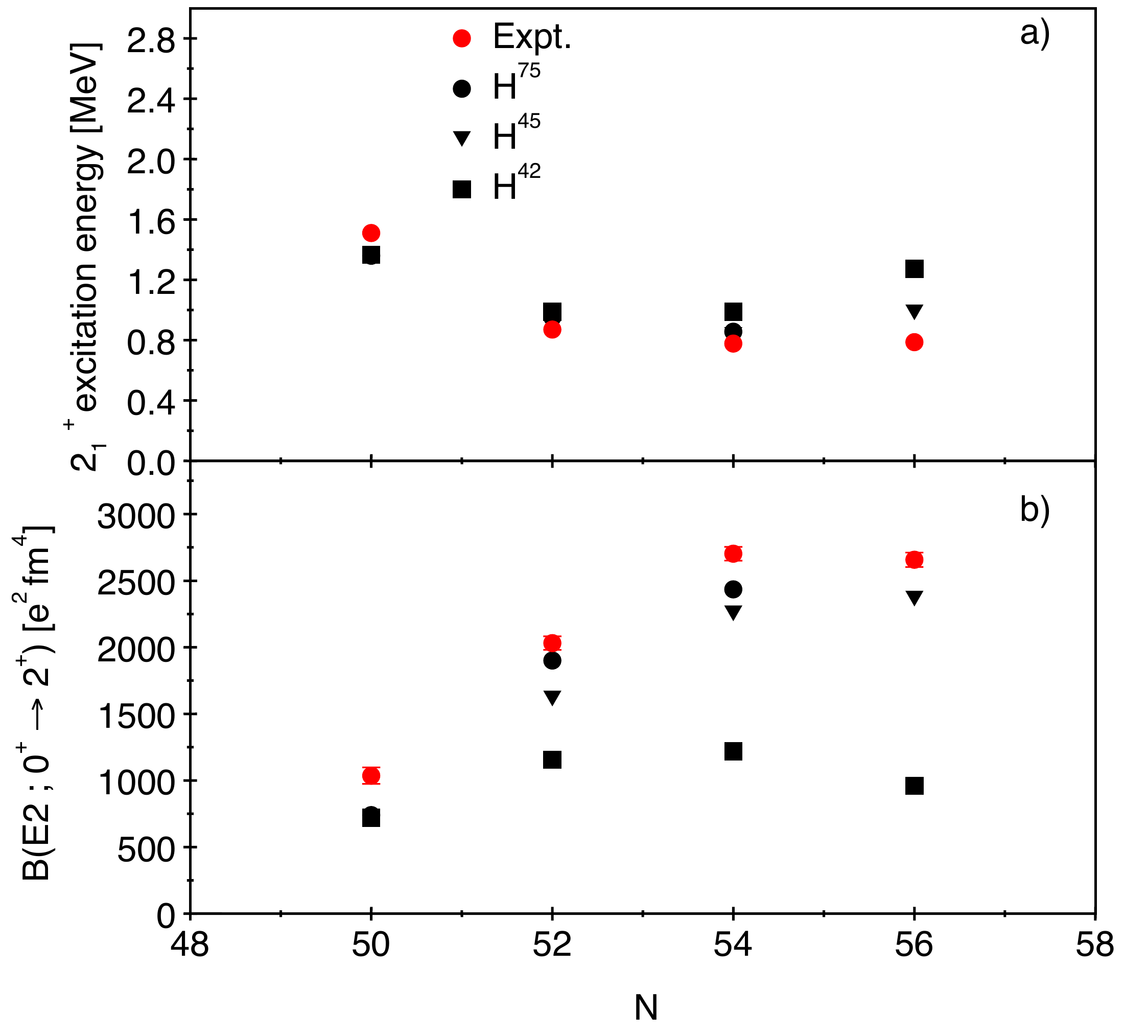}
\caption{(Color online) Same as in Fig. \ref{Zr}, for Mo isotopes.} 
\label{Mo}
\end{center}
\end{figure}

In Fig. \ref{Mo} we show the behavior of the calculated
$E^{ex}_{2^+_1}$ and $B(E2;0^+_1 \rightarrow 2^+_1)$ transition rates
as a function of $N$, and compare them with the available experimental
data.
The results obtained using $H^{75}$ are reported only up to $N=54$,
since from $N=56$ on the dimensions of the eigenvalue problem are out
of reach for the shell-model code we have employed.

\begin{figure}[H]
\begin{center}
\includegraphics[scale=0.37,angle=0]{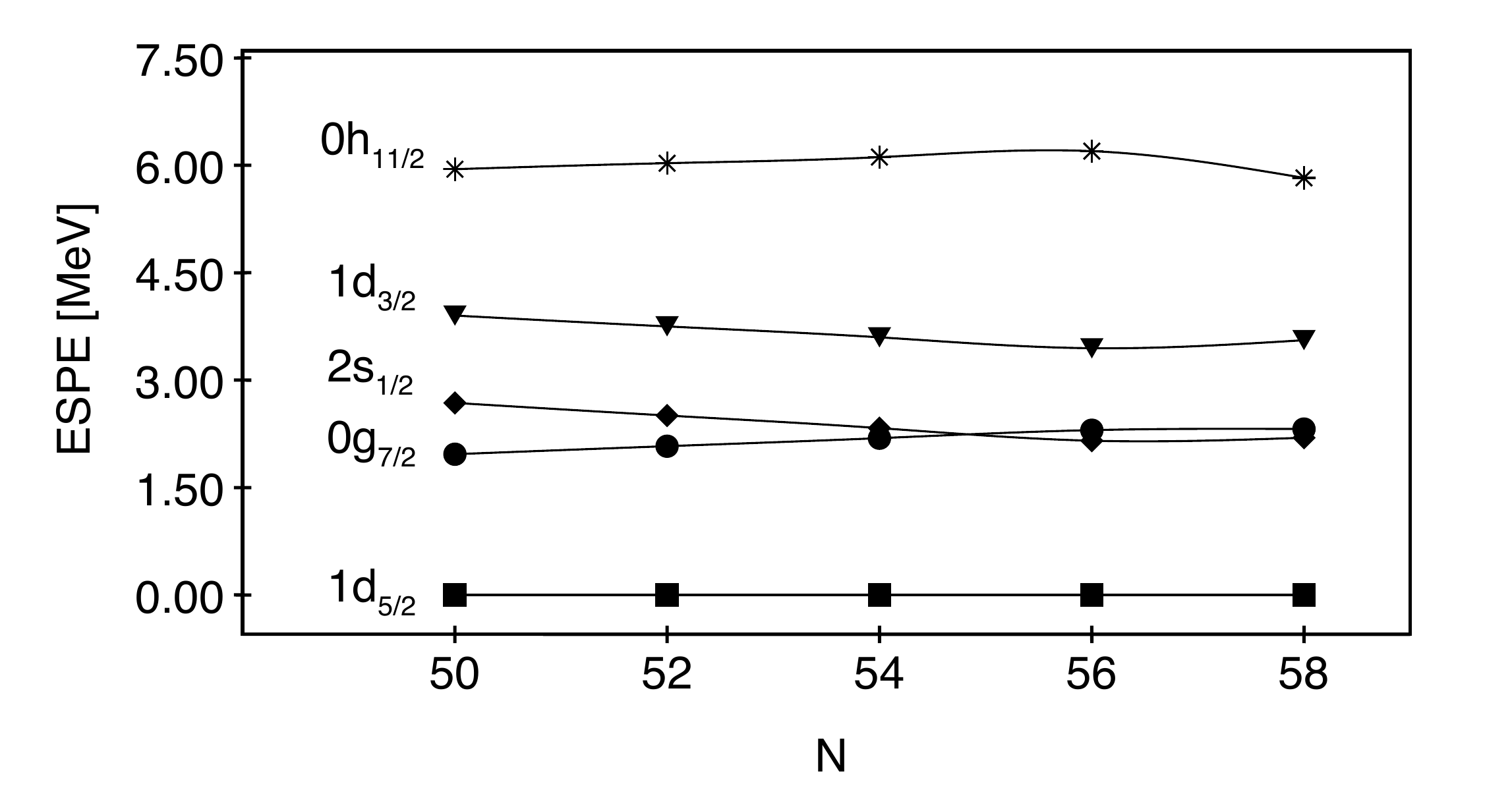}
\caption{Same as in Fig. \ref{espeZr}, for Mo isotopes.}
\label{espeMo}
\end{center}
\end{figure}

Mo isotopes have four valence protons outside $^{88}$Sr,
therefore the $0g_{9/2}$ proton orbital begins to be filled, and
this induces a relevant collectivity by way of the attractive tensor
force acting  between the proton $0g_{9/2}$ and the neutron $0g_{7/2}$
orbitals. 
This implies that the energies are less sensitive to the
single-neutron structure and, consequently, a $N=56$ subshell closure
is not expected, even if the neutron ESPE reported in
Fig. \ref{espeMo} are similar to those for Zr isotopes.
As a matter of fact, while in the ground state (g.s.) wave function of
$^{96}$Zr the $1d_{5/2}$ neutron orbital is $94 \%$ full, its
occupation drops to $74 \%$ in the $^{98}$Mo g.s., owing to the
strong interaction $\pi 0g_{9/2} - \nu 0g_{7/2}$.
As a consequence, a possible shell closure is obviously prevented, and
the reliability of a neutron model-space truncation holds up to
$N=56$, the eigenvalues of the three hamiltonians being very close to
each other.

The situation is quite different for the $E2$ transition rates;
starting from $N=52$ the results obtained with $H_{\rm eff}^{42}$ are far from
those provided by $H^{75}$, $H_{\rm eff}^{45}$, the former transition rates
being strongly lower than the latters.  
The model space spanned only by the $1d_{5/2}$ $0g_{7/2}$ neutron
orbitals is clearly not able to describe the onset of quadrupole
collectivity, and once again we mention the need to include effective
two-body $E2$ operator to cure this deficiency.

In this connection it is worth to stress the reliability of the
$H_{\rm eff}^{45}$ hamiltonian, whose results are quite close to the
$H^{75}$ ones and in a very satisfactory agreement with the
experimental data.
In particular, the reproduction of the $E2$ strengths seems to
indicate that the increase of the number of protons induces a
quenching of the $Z=38$ proton cross-shell excitations.   

\subsection{Ru isotopes}
\begin{figure}[H]
\begin{center}
\includegraphics[scale=0.37,angle=0]{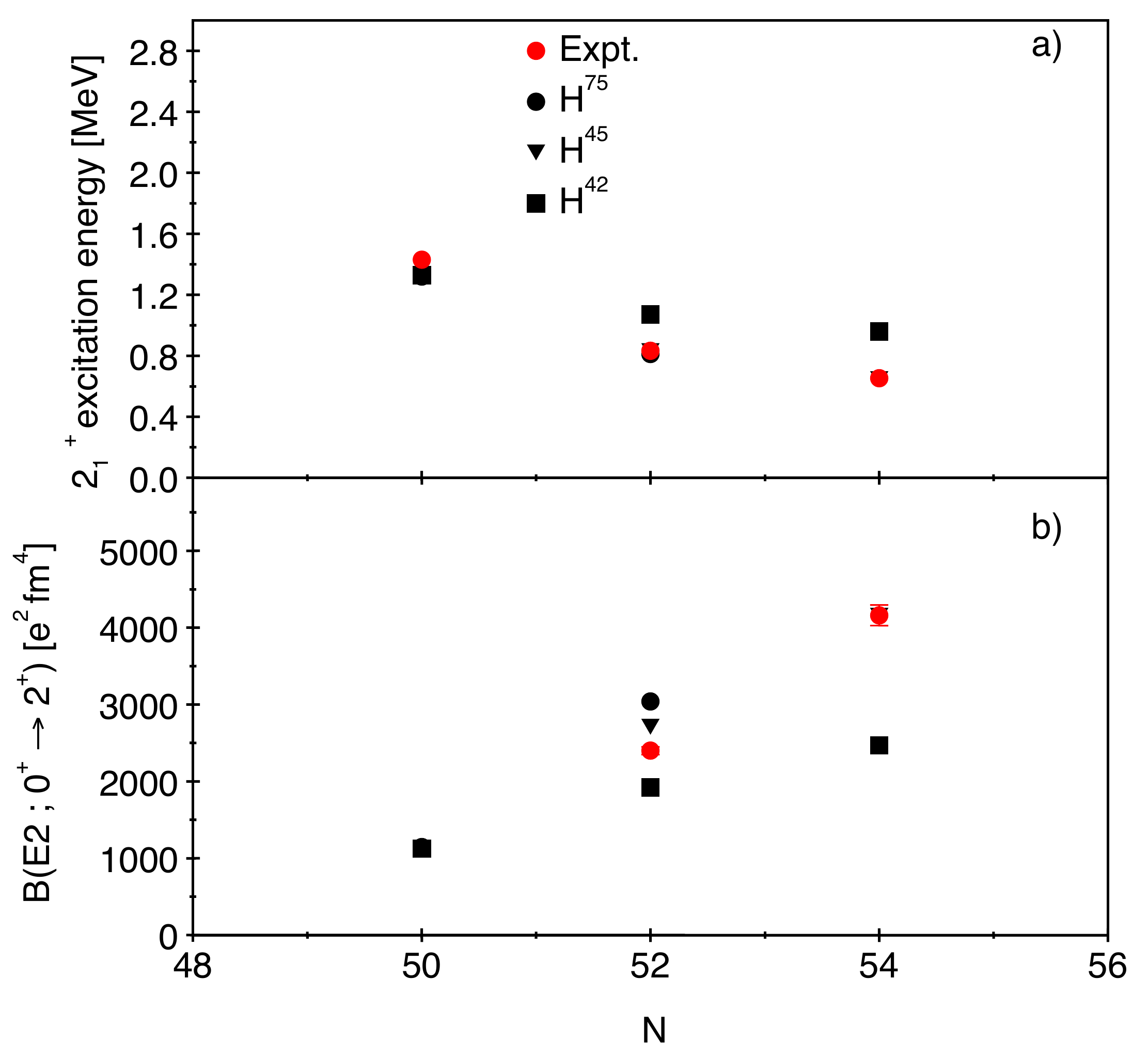}
\caption{(Color online) Same as in Fig. \ref{Zr}, for Ru isotopes.} 
\label{Ru}
\end{center}
\end{figure}

The calculated results for ruthenium isotopes are reported and
compared with the experimental ones in Fig. \ref{Ru}. 
Owing to the increase of the valence proton number, we are able to
perform calculations with $H_{\rm eff}^{45}$  only up to $N=54$, so we cannot
report a comparison between the different truncation schemes for
heavier isotopes.

From the inspection of Fig. \ref{Ru}, we see that predicted energies
with the three hamiltonians are close and in good agreement with the
experiment.
Because of the restricted model space, the $H_{\rm eff}^{42}$ hamiltonian - as for
Mo isotopes - is not able to describe the quadrupole collectivity as
evidenced by the $E2$ transition rates.

The $H_{\rm eff}^{45}$ predicted rates are instead very close to those of the
``mother hamiltonian'', and reproduce nicely the collective behavior.

\begin{figure}[H]
\begin{center}
\includegraphics[scale=0.37,angle=0]{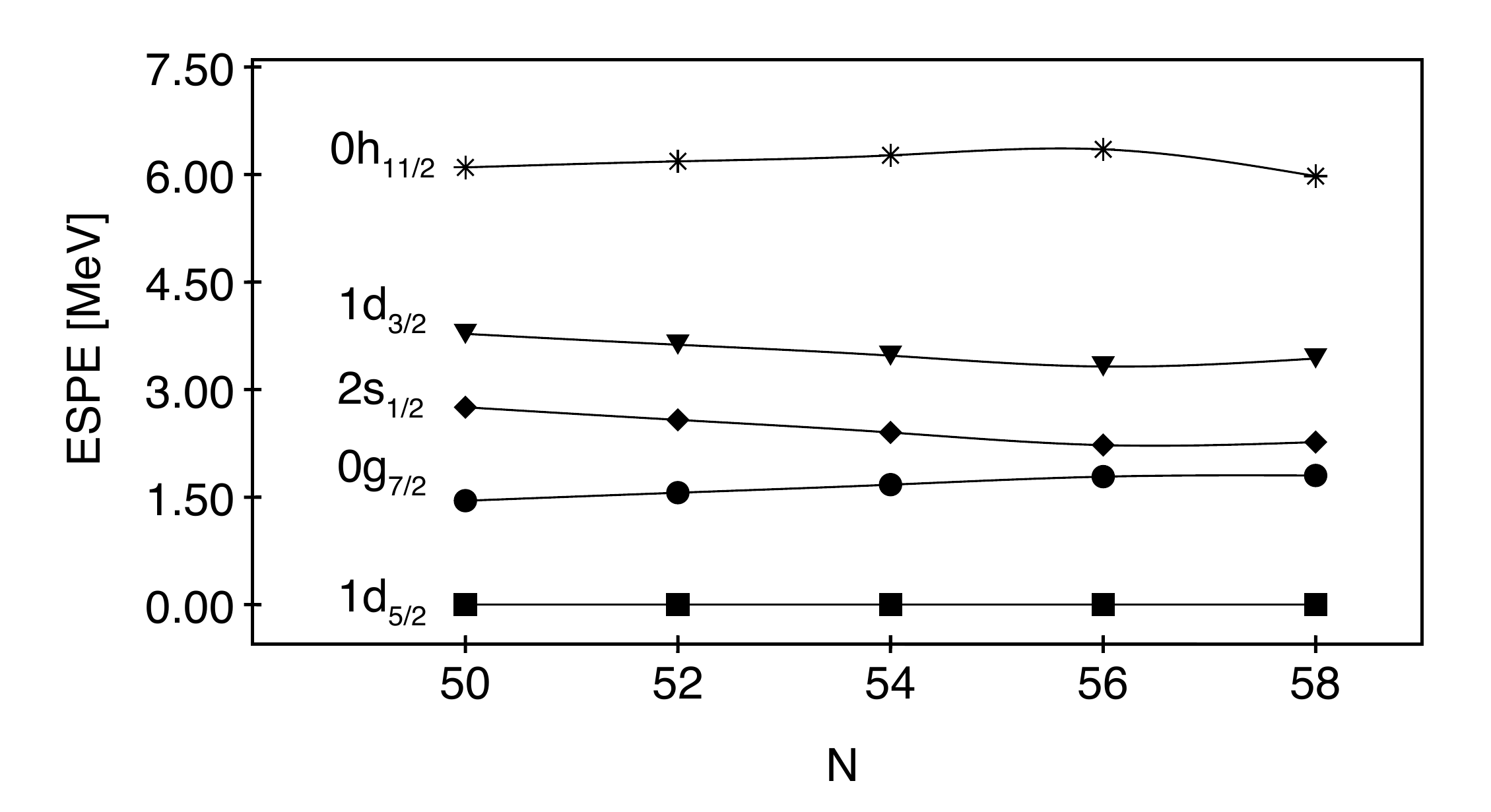}
\caption{Same as in Fig. \ref{espeZr}, for Ru isotopes.}
\label{espeRu}
\end{center}
\end{figure}

In Fig. \ref{espeRu} the calculated neutron ESPE of Ru isotopes
are reported up to $N=58$.
It can be observed a slight lowering of the $0g_{7/2}$ ESPE, the
latter starting to detach from the $1d_{3/2},2s_{1/2}$ ones.

\subsection{Pd isotopes}
\begin{figure}[H]
\begin{center}
\includegraphics[scale=0.37,angle=0]{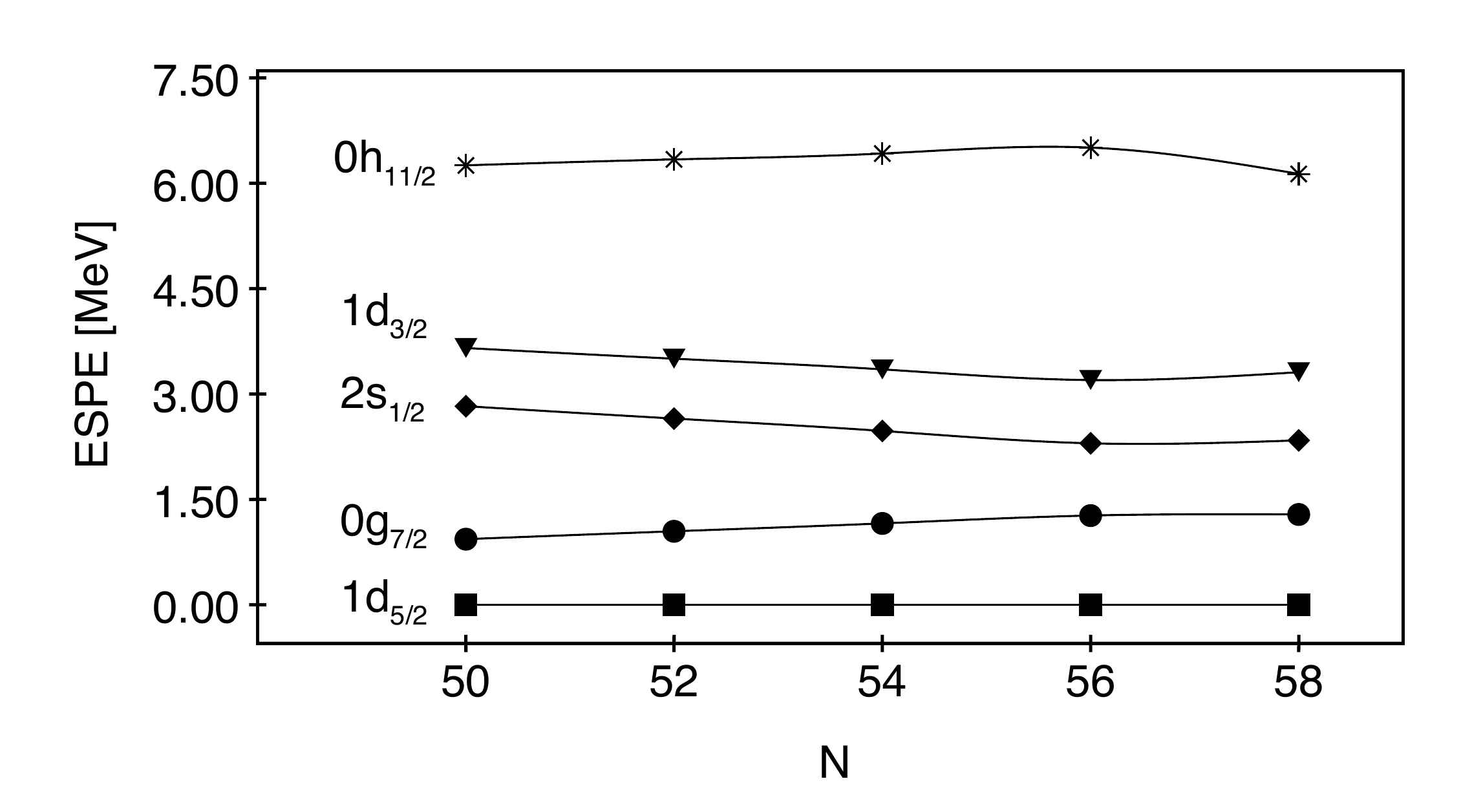}
\caption{Same as in Fig. \ref{espeZr}, for Pd isotopes.}
\label{espePd}
\end{center}
\end{figure}

From the inspection of Fig. \ref{espePd}, where the neutron ESPE for
Pd isotopes are shown as a a function of $N$, it can be seen that the
half filling of the proton $g_{9/2}$ orbital has induced a sizeable
downshift of the neutron $g_{7/2}$ orbital and the rise of an energy
gap ($\simeq 2$MeV) between the $d_{5/2}$ and $g_{7/2}$ orbitals and
the higher ones.

\begin{figure}[H]
\begin{center}
\includegraphics[scale=0.37,angle=0]{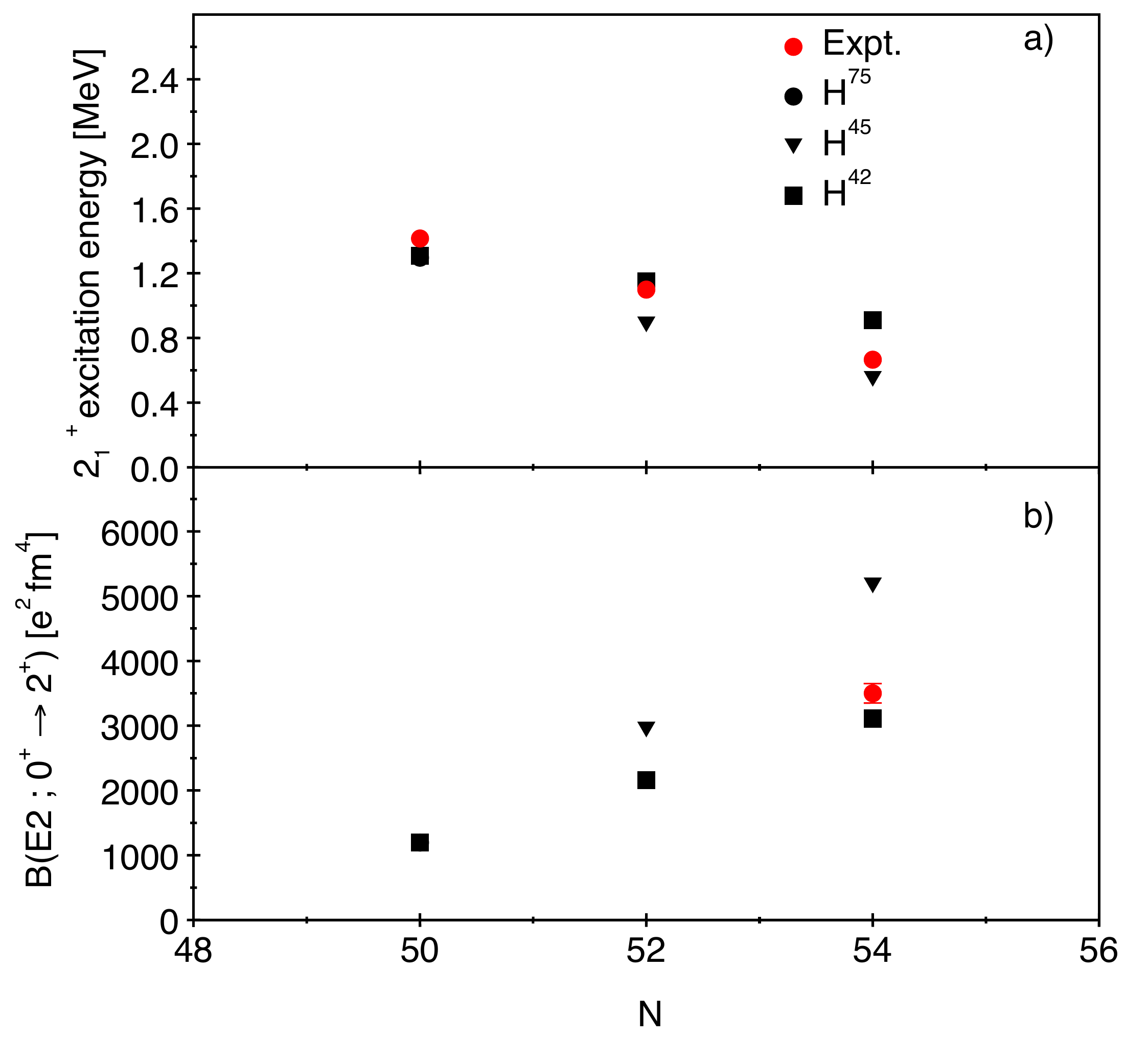}
\caption{(Color online) Same as in Fig. \ref{Zr}, for Pd isotopes.} 
\label{Pd}
\end{center}
\end{figure}

The calculated eigenvalues are in a good agreement with the data up to
$N=54$ (see Fig. \ref{Pd}), $^{100}$Pd being the heaviest computationally attainable
isotope with the $H_{\rm eff}^{45}$ hamiltonian, and the observed behavior of the
$E^{ex}_{2^+_1}$ is satisfactorily reproduced.
The $B(E2;0^+_1 \rightarrow 2^+_1)$ results with $H_{\rm eff}^{45}$ and $H_{\rm eff}^{42}$
are remarkably different, especially for $N=54$. 
This means that the calculated wave functions with $H_{\rm eff}^{45}$ contain
relevant neutron components involving the $s_{1/2}$, $d_{3/2}$ and
$h_{11/2}$ orbitals, testifying the need of an effective two-body $E2$
operator, which will take into account better the neglected neutron
degrees of freedom.

 It is worth to note that the comparison with the sole available
experimental transition rate shows that the $B(E2)$s with $H_{\rm eff}^{45}$
overestimate the quadrupole collectivity, at least for $N=54$.

\subsection{Cd isotopes}

\begin{figure}[H]
\begin{center}
\includegraphics[scale=0.37,angle=0]{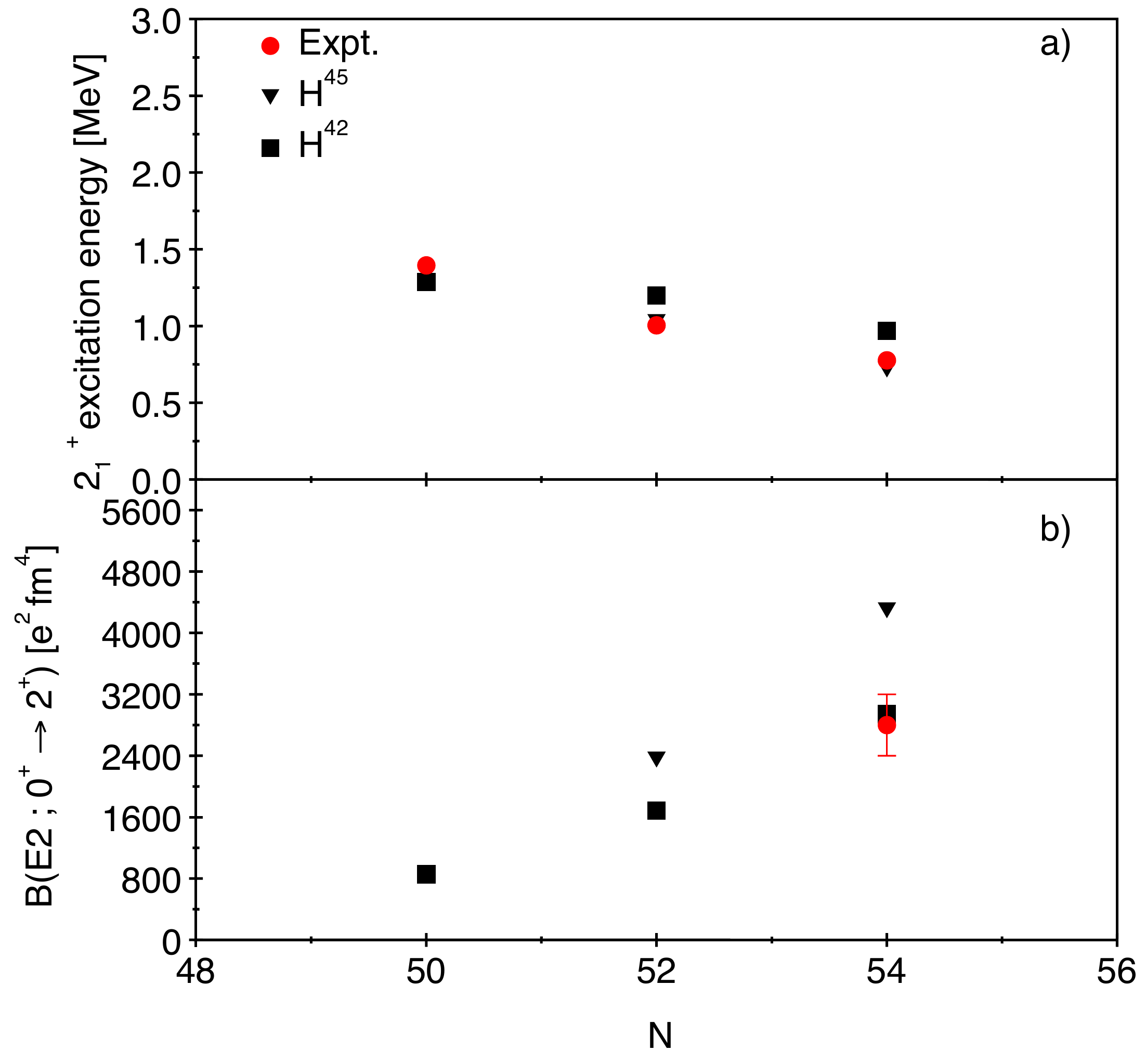}
\caption{(Color online) Same as in Fig. \ref{Zr}, for Cd isotopes.} 
\label{Cd}
\end{center}
\end{figure}

\begin{figure}[H]
\begin{center}
\includegraphics[scale=0.37,angle=0]{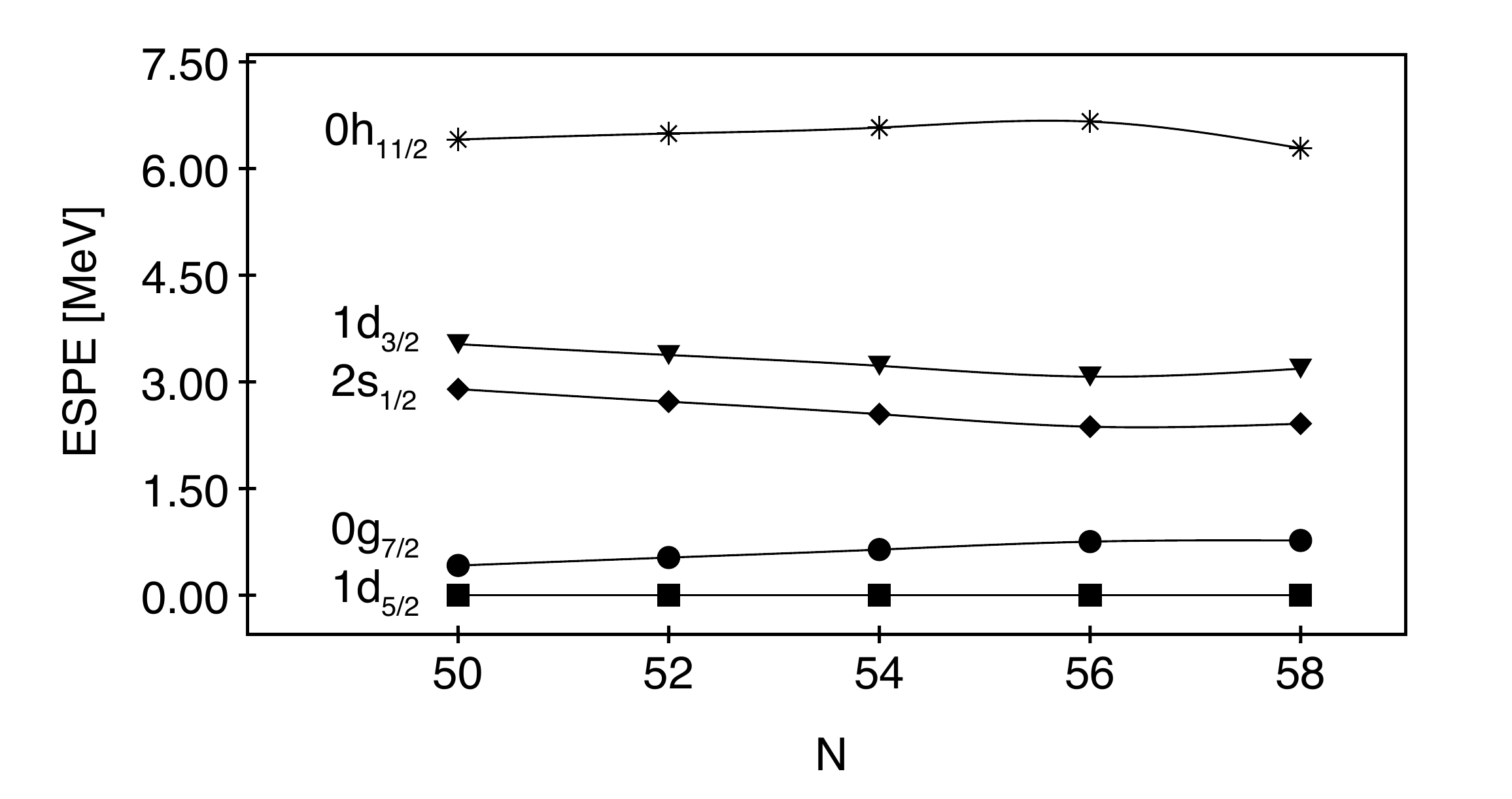}
\caption{Same as in Fig. \ref{espeZr}, for Cd isotopes.}
\label{espeCd}
\end{center}
\end{figure}

In Figs. \ref{Cd} and \ref{espeCd} we report, respectively, the
calculated $E^{ex}_{2^+_1}$ and  $B(E2;0^+_1 \rightarrow 2^+_1)$
transition rates compared with the experimental ones, and the neutron
ESPE.

As it can be seen, almost the same comments made for palladium
isotopes still hold for Cd ones.
The gap between the ($d_{5/2}$, $g_{7/2}$)  and ($s_{1/2}$, $d_{3/2}$
$h_{11/2}$) neutron orbitals increases, due to the progressive
filling of the proton $g_{9/2}$ orbital.
The $H_{\rm eff}^{42}$ hamiltonian is able to reproduce the eigenvalues of the
hamiltonian $H_{\rm eff}^{45}$, while the lack of relevant components
of the wave functions in the subspace spanned by the
$s_{1/2},d_{3/2},h_{11/2}$ neutron orbitals make the calculated
$H_{\rm eff}^{42}$ transition rates for $N=54$ smaller than the
$H_{\rm eff}^{45}$ ones.

\subsection{Sn isotopes}
\begin{figure}[H]
\begin{center}
\includegraphics[scale=0.37,angle=0]{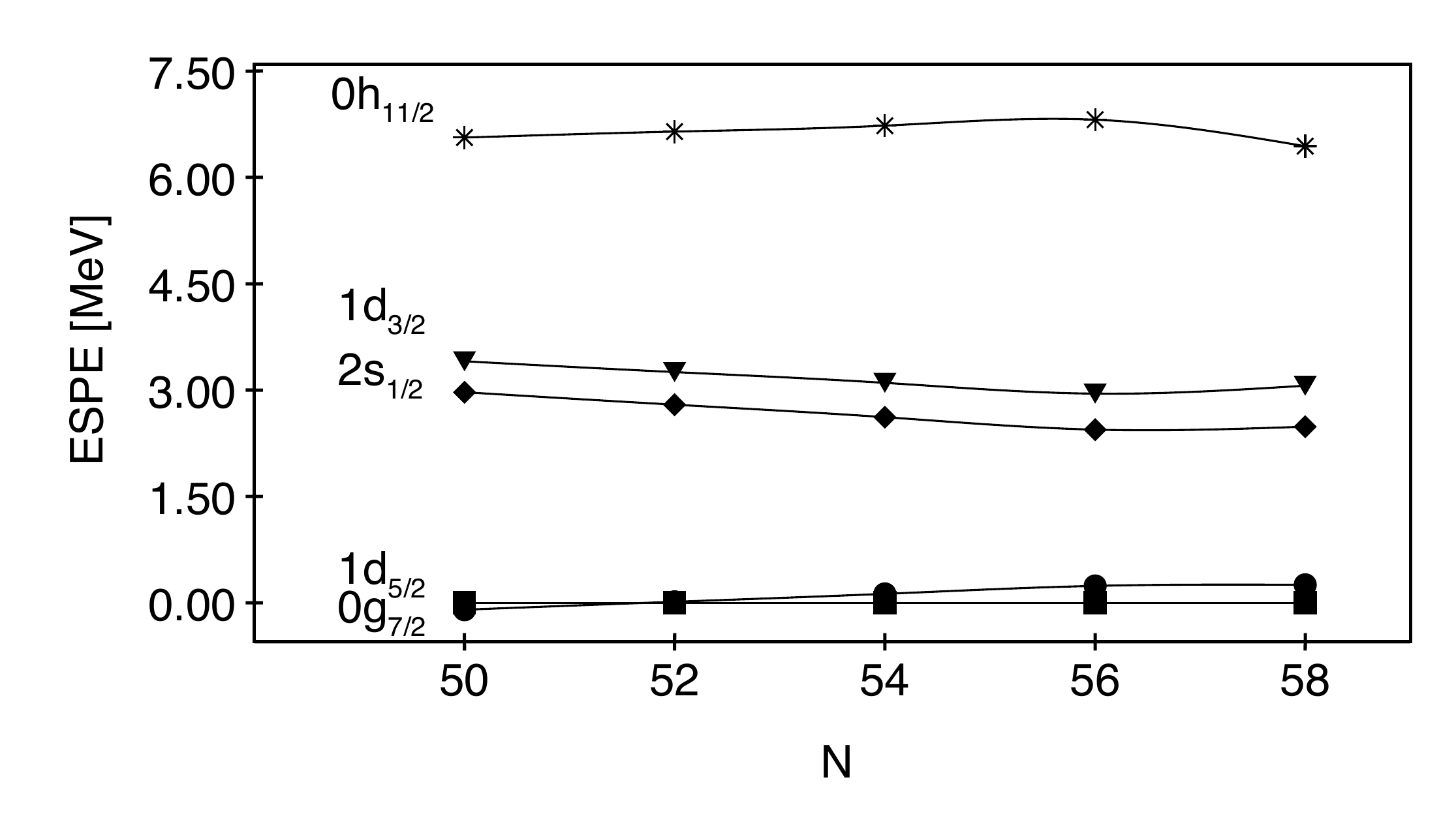}
\caption{Same as in Fig. \ref{espeZr}, for Sn isotopes.}
\label{espeSn}
\end{center}
\end{figure}

The inspection of Fig. \ref{espeSn} evidences that the complete
filling of the proton $0g_{9/2}$ orbital induces a $~3 $ MeV energy gap
at $Z=50$ between the $1d_{5/2},0g_{7/2}$ subspace and that spanned by
the $2s_{1/2},1d_{3/2},0h_{11/2}$ orbitals.

\begin{figure}[H]
\begin{center}
\includegraphics[scale=0.37,angle=0]{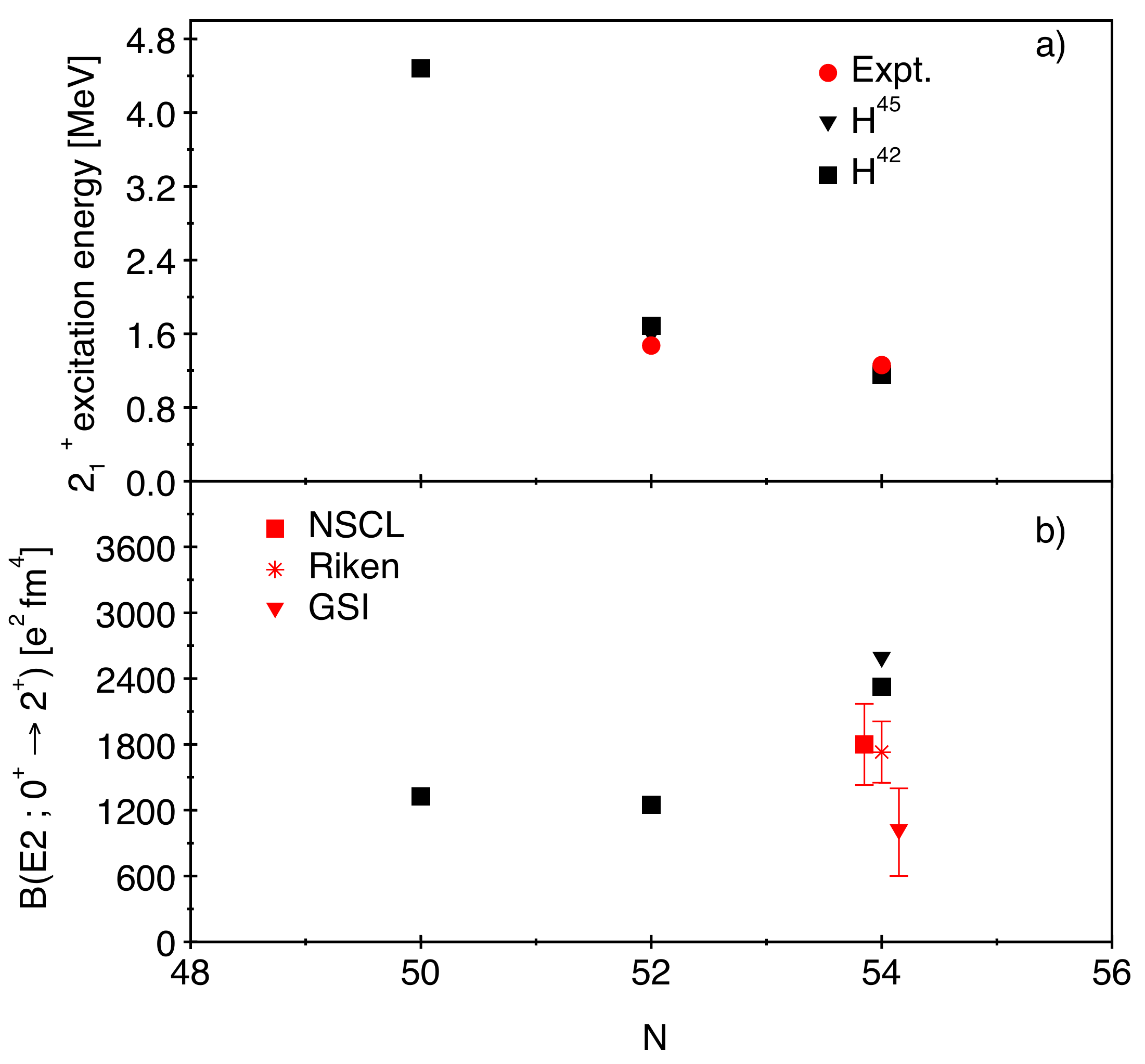}
\caption{(Color online) Same as in Fig. \ref{Zr}, for Sn isotopes.
The experimental transition strengths for $^{104}$Sn are from
Refs. \cite{Guastalla13,Bader13,Doornenbal14}.} 
\label{Sn}
\end{center}
\end{figure}

 This, obviously, suddenly enforces the effectiveness of our truncation
scheme when using $H_{\rm eff}^{42}$.
As a matter of fact, the quadrupole electric transition rates and the
excitation energies of the yrast $2^+$ states calculated with
$H_{\rm eff}^{42}$, reported in Fig. \ref{Sn},  are in good agreement with those
obtained with $H_{\rm eff}^{45}$.
Moreover, it can be seen that the calculations with $H_{\rm eff}^{42}$ are able to
reproduce quite well the experimental $E^{ex}_{2^+_1}$ and the
$B(E2;0_1^+ \rightarrow 2^+_1)$ up to $N=54$ and, consequently, the
onset of collectivity from $^{102}$Sn on that is driven by the $Z=50$
cross-shell excitations (see Ref. \cite{Coraggio15a} for a more
complete discussion about this point). 

Before concluding this subsection, it is in order to point out that
the calculations for $^{104}$Sn with the $H_{\rm eff}^{45}$
hamiltonian have been performed limiting to 6 the maximum number $t$
of valence particles allowed in the proton $1d_{5/2},0g_ {7/2}$ and
neutron $2s_{1/2},1d_{3/2},0h_{11/2}$ orbitals, because of the
limitations of the ANTOINE shell-model code.
To check the reliability of such truncation, we have studied the
behavior of the calculated quantities as a function of $t$.
The obtained results are quite stable from $t=4$ on, the
$E^{ex}_{2^+_1}$ and the $B(E2;0_1^+ \rightarrow 2^+_1)$ varying less
than $1\%$ and $2\%$ respectively, passing from $t=5$ to $t=6$.

\section{Summary and outlook}

The aim of the present paper has been to introduce an original
double-step approach to simplify the computational problem of
large-scale shell-model calculations.
The core of our method is the study of the ESPE of the large-scale
hamiltonian, so as to identify the most relevant degrees of freedom to
be taken into account in the construction of a truncated shell-model
hamiltonian. 
To this end, a unitary transformation is employed so to derive new
effective shell-model hamiltonians defined within a reduced set of
single-particle orbitals, accordingly to the ESPE analysis.

We have applied this procedure to a realistic shell-model hamiltonian,
whose model space was designed so to describe the $Z=50$ cross-shell
excitations for nuclei outside $^{88}$Sr employing seven proton and
five neutron orbitals.
Because of the behavior of the proton and neutron ESPE, we have
identified two new model subspaces made up by four proton orbitals and
five and two neutron ones, respectively, and transformed our original
hamiltonian in these subsets.

We have performed calculations for Zr, Mo, Ru, Pd, Cd, and Sn isotonic
chains to check the reliability of our procedure.
The results obtained with the effectively truncated
hamiltonians indicates the ability to reproduce eigenvalues and
electromagnetic transition rates of the original shell-model
hamiltonian, especially when the ESPE provide a neat separation in
energy between the new model subspaces and their complement.

This double-step approach may provide a reliable truncation procedure in any
large-scale shell-model calculation.
As a matter of fact, we are intended to apply it in other regions, as
for example to study isotopic and isotonic chains with valence particle
outside doubly-magic $^{132}$Sn and $^{208}$Pb, where the large model
spaces leads to critical situations of the computational complexity,
when increasing the number of valence nucleons.

We are also studying the possibility to extend this procedure so to
truncate shell-model spaces freezing the degrees of freedom of filled
nuclear orbitals, as we already experienced in Ref. \cite{Coraggio14b}
but within a perturbative derivation of the reduced effective
hamiltonians, and to calculate effective two-body
electromagnetic operators, consistently with the unitary
transformation of the hamiltonian, so to improve also the reproduction
of the corresponding transition rates.

\bibliographystyle{apsrev}
\bibliography{biblio.bib}

\end{document}